\pdfoutput=1

\documentclass[11pt]{article}

\usepackage[final]{acl}

\usepackage{times}
\usepackage{latexsym,multirow,multicol,makecell,booktabs,enumerate,stfloats,url}

\usepackage[T1]{fontenc}

\usepackage[utf8]{inputenc}

\usepackage{microtype}

\usepackage{inconsolata}

\usepackage{graphicx}

\title{QualiSpeech: A Speech Quality Assessment Dataset  with \\ Natural Language Reasoning and Descriptions}

\author{Siyin Wang$^1$, Wenyi Yu$^1$, Xianzhao Chen$^2$,   Xiaohai Tian$^2$, Jun Zhang$^2$ \\ \textbf{Lu Lu}$^2$, \textbf{Yu Tsao}$^3$, \textbf{Junichi Yamagishi}$^4$, \textbf{Yuxuan Wang}$^2$, \textbf{Chao Zhang}$^{1}$\thanks{$^{}$Corresponding author.} \\
  $^1$Tsinghua University, $^2$ByteDance, $^3$Academia Sinica, $^4$National Institute of Informatics \\
  \texttt{wangsiyi23@mails.tsinghua.edu.cn,cz277@tsinghua.edu.cn} \\}

\begin{document}
\maketitle
\begin{abstract}
This paper explores a novel perspective to speech quality assessment by leveraging natural language descriptions, offering richer, more nuanced insights than traditional numerical scoring methods. Natural language feedback provides instructive recommendations and detailed evaluations, yet existing datasets lack the comprehensive annotations needed for this approach. To bridge this gap, we introduce \textbf{QualiSpeech}, a comprehensive low-level speech quality assessment dataset encompassing 11 key aspects and detailed natural language comments that include reasoning and contextual insights. Additionally, we propose the QualiSpeech Benchmark to evaluate the low-level speech understanding capabilities of auditory large language models (LLMs). Experimental results demonstrate that finetuned auditory LLMs can reliably generate detailed descriptions of noise and distortion, effectively identifying their types and temporal characteristics. The results further highlight the potential for incorporating reasoning to enhance the accuracy and reliability of quality assessments. The dataset can be found at \url{https://huggingface.co/datasets/tsinghua-ee/QualiSpeech}.

\end{abstract}

\section{Introduction}

\begin{figure}[t]
\hspace{0.02cm}
\includegraphics[width=0.5\textwidth]{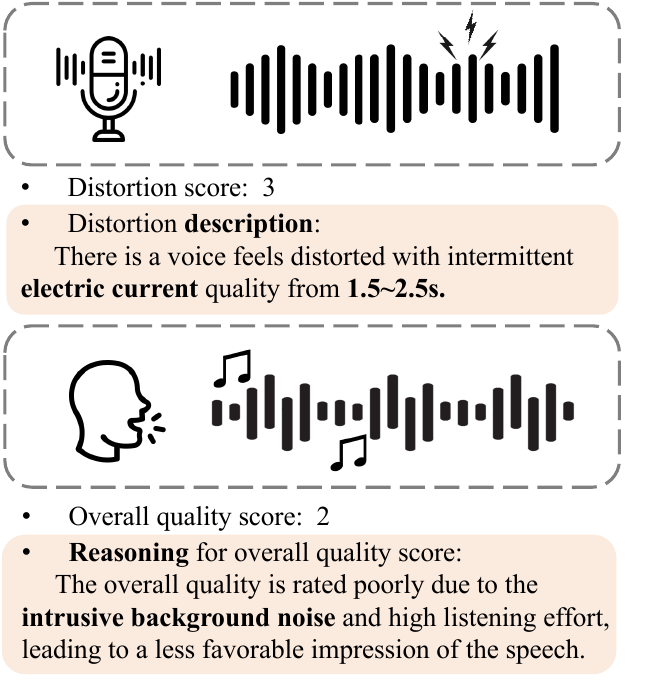}
\caption{Examples from QualiSpeech. QualiSpeech is a comprehensive low-level speech quality assessment dataset that includes numerical scores for 7 aspects, specific descriptions for 4 aspects, and a detailed natural language comment on overall quality, incorporating reasoning and contextual insights. The examples highlight a simple distortion description alongside the reasoning section from a descriptive comment. 
}
\label{fig1}
\vspace{-0.4cm}
\end{figure}

Assessing speech quality is essential for evaluating the performance of speech synthesis systems and identifying distortions in communication networks \cite{bvcc,nisqa}. The gold standard for speech quality evaluation remains human assessment, typically measured using the mean opinion score (MOS) — an average rating derived from listening tests that gauge overall audio quality \cite{scale}. However, human evaluations are both time-intensive and laborious, prompting the development of automated methods for speech quality assessment. These methods, often powered by deep learning, enable rapid and scalable evaluation across large datasets \cite{automos,mosnet}. Most current approaches focus on MOS prediction, generating a numerical score that represents the perceived quality of the speech \cite{ssl-mos,utmos}. While these scores allow for straightforward comparisons between samples, they do not reveal the reasoning behind a particular score, leaving the underlying quality factors unexplained.


Evaluating speech quality using natural language offers a novel and intuitive approach, enabling more nuanced and detailed feedback compared to traditional numerical scales. Descriptive evaluations analyze multiple low-level speech features and synthesize them into an overall assessment, offering instructive insights for applications such as improving speech synthesis systems. For instance, a description like ``\textit{There is a voice that feels distorted with intermittent electric current quality from 1.5$\sim$2.5s}'' provides richer context than a standalone distortion score, as shown in Figure \ref{fig1}. To address the absence of datasets tailored for natural language-based speech quality assessment, we present \textbf{QualiSpeech}, the first dataset designed to capture diverse low-level speech characteristics through detailed descriptive comments.


Recent advancements in auditory LLMs have made natural language-based speech quality evaluation increasingly feasible \cite{salmonn,qwenaudio,audiopalm}. By integrating speech encoders with powerful LLM backbones, these models excel in high-level spoken language understanding tasks. However, low-level speech perception tasks remain largely underexplored in both the training and evaluation of current auditory LLMs \cite{dynamicsuperb,airbench}. To address this limitation, we propose the QualiSpeech benchmark on multi-choice speech quality assessment tasks, revealing that existing auditory LLMs struggle to assess speech quality accurately. Leveraging the QualiSpeech dataset, we enhance the SALMONN-7B \cite{salmonn} model with the ability to provide natural language descriptions of speech quality across multiple dimensions. This model demonstrates the ability to produce detailed and precise descriptions of noise and distortion, underscoring the advantages of using natural language for nuanced speech quality evaluation. We also demonstrate the feasibility of reasoning for speech quality assessment using text LLMs.

Our contributions can be summarized as follows:

\begin{itemize}
    \item We introduce QualiSpeech, a database for speech quality assessment using natural language descriptions. QualiSpeech evaluates speech of both humans and various text-to-speech (TTS) synthesis systems across a comprehensive range of aspects, encompassing diverse artificial distortions and real-world scenarios. To the best of our knowledge, QualiSpeech is the first dataset designed using low-level speech perception annotated with detailed natural language descriptions for quality assessment of both synthetic and real speech.
    \item We also develop an auditory LLM that can assess speech quality across multiple aspects by generating detailed descriptive comments using the QualiSpeech dataset.
    \item We propose QualiSpeech benchmark for low-level speech understanding ability evaluation of auditory LLMs.
\end{itemize}

\section{Related Work}

\subsection{Speech quality assessment dataset}

To address the need for automatic speech quality assessment models for evaluating speech synthesis systems, the BVCC \cite{bvcc} dataset is proposed by collecting speech samples from past Blizzard Challenges \cite{blizzard} and VCC Challenges \cite{vcc} with a standardized new MOS score. More speech assessment datasets for evaluating speech synthesis models are introduced to extend application scenarios \cite{voicemos2023} and mitigate the influence of speaker \cite{somos}. From a different perspective, the NISQA dataset \cite{nisqa} focuses on real-world speech recordings and simulated distortions commonly found in communication networks.


Before the introduction of QualiSpeech, prior research works have typically treated the evaluation of synthetic and real speech as separate tasks due to their distinct characteristics. Synthetic speech is typically free of noise but often lacks naturalness, while real speech is more affected by noise than by issues of naturalness. By providing detailed aspect scores and reasoning in descriptive comments, QualiSpeech aims to facilitate the development of general speech quality assessment models capable of effectively distinguishing between these different types of speech.



Previous methods for automatic speech quality assessment have predominantly focused on MOS prediction. Deep learning-based approaches can achieve strong correlations with MOS derived from human evaluations \cite{automos,mosnet,ssl-mos}. Expanding beyond overall scores, NISQA \cite{nisqa} introduced score-based prediction of specific speech quality dimensions, including noise, colouration, discontinuity, and loudness. Building on this foundation, QualiSpeech seeks to push the field further by enabling the development of more advanced speech quality assessment models. These models are designed to analyze a wider range of low-level speech features while providing detailed reasoning in natural language.

\begin{figure}[t]
\hspace{0.22cm}
\includegraphics[width=2.8in]{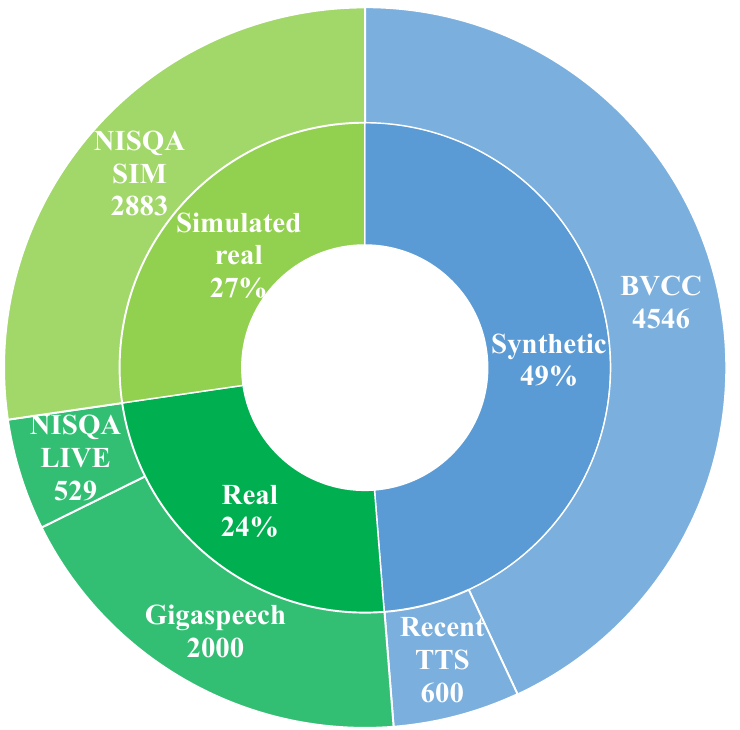}
\caption{The dataset source of train split of QualiSpeech dataset. It has a balanced distribution of synthetic data and Real data (including simulated real data). The total size of the train split is 10,558.}
\label{fig2}
\vspace{-0.4cm}
\end{figure}

\subsection{Auditory LLM for speech perception}

Auditory LLMs \cite{speechllm,salmonn,qwenaudio,wavllm,DeSTA,speechcraft} have shown remarkable performance across a broad spectrum of high-level speech perception tasks, including speech recognition \cite{yassir,wenyi}, translation \cite{wujian,salm}, understanding \cite{slu}, and speaker and emotion recognition \cite{WuZhiYong}. By harnessing the power of LLMs to analyze speech data and generate natural language responses, auditory LLMs hold significant potential to unify diverse speech understanding tasks within a single framework \cite{salmonn,qwen2audio}. To comprehensively evaluate these capabilities, benchmarks such as Dynamic-Superb \cite{dynamicsuperb}, AIR-Bench \cite{airbench}, and AudioBench \cite{audiobench} have been introduced, offering valuable insights into the strengths and limitations of auditory LLMs. However, low-level speech quality assessment tasks remain largely neglected. To fill this gap, we propose a multi-choice benchmark specifically tailored for low-level speech perception tasks, utilizing scores from seven key dimensions in QualiSpeech.

Recent efforts have expanded the range of tasks auditory LLMs can tackle, including spatial audio processing \cite{bat,tang2024can} and audio entailment \cite{audioentail}. The potential of using auditory LLMs for speech quality assessment has also been explored in \cite{llm1,llm2,chenchen}. However, these efforts remain largely focused on MOS prediction, which fails to fully leverage the unique capability of auditory LLMs to generate rich, natural language responses. By utilizing the diverse annotations in QualiSpeech, we can develop an auditory LLM for speech quality assessment that offers a more holistic and detailed evaluation, providing nuanced descriptions of speech quality rather than relying solely on numerical scores.

\section{QualiSpeech Dataset and Benchmark}

\subsection{Dataset}


We introduce QualiSpeech, an English-language speech quality assessment dataset designed to encompass diverse scenarios and aspects. It provides the most comprehensive set of low-level speech feature annotations available to date, including numerical scores across seven dimensions and concise descriptions for four aspects. Significantly, QualiSpeech pioneers the use of natural language descriptions, offering detailed and logically structured assessments that go beyond traditional numerical scoring methods.

\begin{figure*}[t]
\centering
\includegraphics[width=\textwidth]{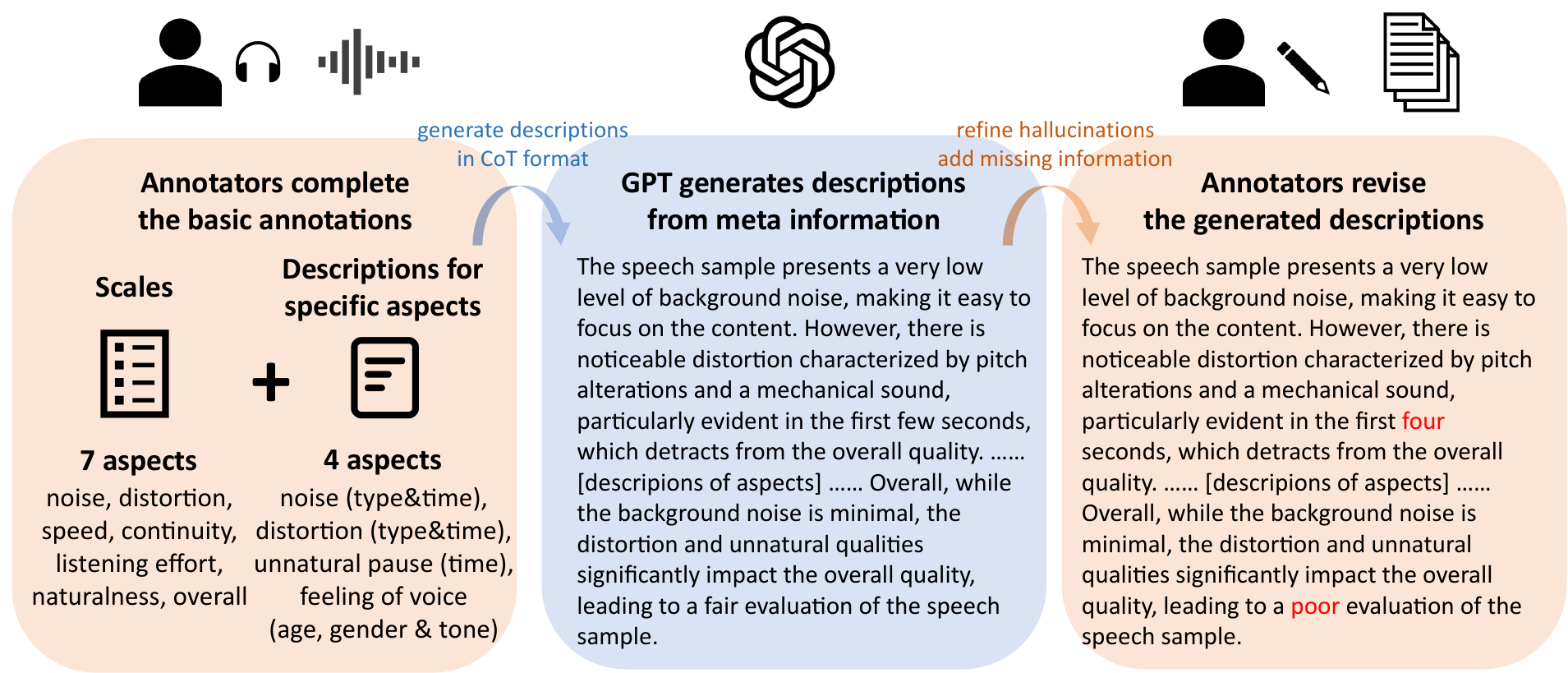}
\vspace{-0.1cm}
\caption{The annotation process of QualiSpeech dataset. In step 1, Listeners annotate basic low-level speech perception characteristics including 7 scores and 4 specific descriptions. In step 2, GPT generates natural language descriptions from annotated meta information. In step 3, annotators check and revise the generated descriptions.}
\label{fig3}
\end{figure*}

\subsubsection{Data collection}

We constructed the QualiSpeech dataset using a diverse range of sources. Figure \ref{fig2} illustrates the data sources for the training split, with detailed statistics provided in the Appendix \ref{append:b}. For synthetic speech, we utilized the BVCC dataset, which includes samples of varying quality generated by diverse systems from past Blizzard and VCC Challenges. To maintain consistency, samples shorter than two seconds were excluded. However, as BVCC was collected in 2021, it lacks data from more recent TTS models. To address this gap, we incorporated synthetic speech generated by 10 recent open-source TTS models, including ChatTTS\footnote{\url{https://github.com/2noise/ChatTTS}}, XTTS v2\footnote{\url{https://github.com/coqui-ai/TTS}}, CosyVoice\footnote{\url{https://github.com/FunAudioLLM/CosyVoice}} \cite{cosyvoice}\, F5-TTS\footnote{\url{https://github.com/SWivid/F5-TTS}\label{f5tts}} \cite{f5tts}, E2 TTS (implemented by F5-TTS\textsuperscript{\ref{f5tts}}) \cite{e2tts}, OpenVoice V1\footnote{\url{https://github.com/myshell-ai/OpenVoice}\label{openvoice}}, OpenVoice V2\textsuperscript{\ref{openvoice}} \cite{openvoice}, Parler-TTS Mini\footnote{\url{https://github.com/huggingface/parler-tts}\label{parlertts}}, Parler-TTS Large\textsuperscript{\ref{parlertts}} and VoiceCraft-830M\footnote{\url{https://github.com/jasonppy/VoiceCraft}} \cite{voicecraft}. Sentences for synthesis were sourced from the SOMOS sentence corpus \cite{somos}, which spans 10 domains, including conversational dialogue, news, and Wikipedia entries. Each TTS model generated 72 samples, distributed as 60 for training, 6 for validation, and 6 for testing. For zero-shot TTS models (except Parler-TTS), prompt audio determining the speaker of the synthesized speech was sampled from LibriHeavy \cite{libriheavy}. For Instruct-TTS Parler-TTS, text descriptions specifying speakers were either sourced from available built-in speakers or generated by GPT. The full list of text descriptions is provided in the Appendix \ref{append:e}.

Speech synthesis models are typically trained exclusively on clean speech, making them unlikely to produce samples that are both unnatural and noisy. To address this gap, 20\% of the synthetic data is mixed with noise. The noise files are sourced from the DNS Challenge dataset \cite{dns}, which draws its content from AudioSet \cite{audioset}, Freesound \cite{freesound}, and DEMAND \cite{demand}. The signal-to-noise ratio is uniformly sampled within the range of 0 to 15.


For real speech, NISQA \cite{nisqa} is exploited due to it has speech samples in rich simulated and live conditions. The simulated distortions are designed to replicate real-world transmission channels. Specifically, the simulated datasets (NISQA\_TRAIN\_SIM and NISQA\_VAL\_SIM) include 1,258 unique distortions generated by combining 9 basic distortion types. From these, one out of every five distortions is selected, resulting in 2,883 samples for the training split and 826 for the validation split. For live communication recordings, we include Skype recordings from NISQA\_TRAIN\_LIVE and NISQA\_VAL\_LIVE. NISQA\_TEST\_FOR and NISQA\_TEST\_P501, the two test sets of NISQA featuring both real and simulated conditions, are incorporated into the test split. We also source real speech from GigaSpeech \cite{gigaspeech}, which includes audiobooks, podcasts, and YouTube recordings. Using UTMOS \cite{utmos}, MOS scores are predicted to classify GigaSpeech S samples into four quality groups. To ensure balanced representation across varying speech qualities, an equal number of samples is selected from each group.

\begin{figure*}[t]
\centering
\includegraphics[width=\textwidth]{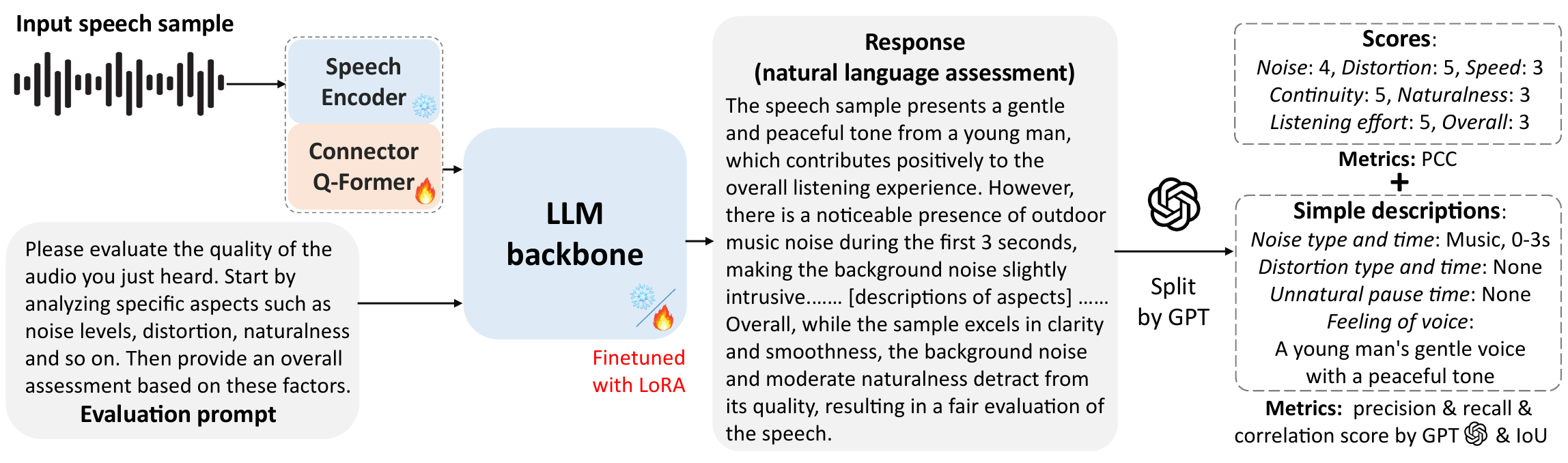}
\vspace{-0.1cm}
\caption{Finetuning auditory LLMs on QualiSpeech and the evaluation procedure of natural language speech assessment. The generated assessment will be split into scores and specific descriptions using GPT first, then PCC is calculated for scores and the correlation score generated by GPT will be used to evaluate the correctness of specific descriptions.}
\label{fig5}
\end{figure*}

\subsubsection{Annotation process}

The annotation process, illustrated in Figure \ref{fig3}, comprises three main steps. The first step involves collecting detailed low-level speech features through listening tests, which include numerical scores across seven aspects and concise descriptions for four specific aspects. Listeners are tasked with rating various low-level features of each speech sample on a five-point scale, where higher scores indicate better quality. The annotated aspects include noise, distortion, speed, continuity, listening effort, naturalness, and overall quality. Our scoring scales are designed following the guidelines in \cite{scale}, with detailed scales and instructions provided in Figure \ref{fig10}. Additionally, the listening tests capture specific descriptions of noise (type and occurrence time), distortion (type and occurrence time), unnatural pauses (occurrence time), and vocal characteristics (perceived age, gender, and tone). These annotations aim to facilitate the generation of more detailed and context-rich natural language descriptions.


The second step is to generate natural language descriptions using GPT\footnote{The version is gpt-4o-mini-2024-07-18.} from the annotated meta information. All annotated aspects are provided to GPT to produce descriptions in a chain-of-thought (CoT) format. This approach ensures the descriptions first analyze low-level speech features before concluding with an overall quality assessment. The third step focuses on refining the generated descriptions. Since GPT can produce hallucinations or inaccuracies, this step is essential for ensuring high-quality natural language descriptions \cite{depictqa}. Annotators review and correct errors, such as inconsistencies between the descriptions and annotations or unsupported claims. They are also tasked with adding any missing aspects to ensure the descriptions fully capture all annotated information. Furthermore, annotators refine the reasoning behind the overall quality assessment, improving the logical coherence and clarity of the descriptions as necessary.

\subsection{Benchmark}

We also establish the QualiSpeech benchmark to assess the low-level speech perception capabilities of auditory LLMs. This multi-choice benchmark spans seven aspects of low-level speech understanding: noise, distortion, speed, continuity, listening effort, naturalness, and overall quality. It is built on the numerical scoring scales provided in the QualiSpeech dataset. Auditory LLMs are tasked with selecting the most appropriate score for a given speech sample on a specific low-level aspect, with the relevant scale provided as guidance. Open-ended question-answering is excluded from the benchmark, as current LLMs often struggle to reliably follow instructions they have not encountered before.

\begin{table*}[t]
\setlength{\tabcolsep}{3pt}
\centering
\begin{tabular}{c|ccccccc}
\toprule
\multirow{2}{*}{Model} & \multicolumn{7}{c}{Aspect of low-level speech perception} \\
& Noise & Distortion & Speed & Continuity & Effort & Naturalness & Overall \\
\midrule
SALMONN-7B & 0.003 & 0.013 & 0.001 & nan & nan & 0.030 & 0.084 \\
SALMONN-13B & 0.001 & 0.002 & 0.025 & -0.001 & -0.069 & 0.013 & 0.100 \\
Qwen-Audio-Chat & \textbf{0.014} & -0.003 & 0.017 & 0.145 & \textbf{0.150} & \textbf{0.148} & \textbf{0.250}\\
Qwen2-Audio-7B-Instruct & -0.048 & \textbf{0.056} & \textbf{0.111} & \textbf{0.201} & 0.035 & -0.082 & 0.112\\
WavLLM & -0.021 & -0.069 & 0.003 & -0.001 & -0.007 & 0.005 & 0.071 \\
\bottomrule
\end{tabular}
\vspace{-0.1cm}
\caption{Open-source auditory LLMs on QualiSpeech benchmark. PCC is reported. Some correlation scores are nan because all predicted scores are the same for that aspect.}
\vspace{-0.1cm}
\label{t1}
\end{table*}

\subsection{Evaluation metrics}

For QualiSpeech benchmark evaluation, PCC (Pearson Correlation Coefficient) is used to evaluate the accuracy of predicted scores.

For QualiSpeech dataset evaluation, PCC is also employed to assess aspects represented by numerical values. For aspects in descriptions, we utilize 4 metrics to cover different evaluation dimensions. First, precision and recall are reported to assess the model's ability to accurately determine the presence of noise (or distortion, unnatural pause). From a complementary point of understanding, correlation scores generated by GPT and intersection over union (IoU) scores are presented. The correlation score gauges the overall relevance of the model's descriptions, while the IoU score measures the accuracy of the predicted time intervals. Note that we calculate these two scores only when the model successfully identify the noise (or distortion, natural part). This ensures that these scores will not be influenced by recall since undetected samples will receive a correlation or IoU score of 0. Precision and recall reflect holistic comprehension, while correlation and IoU focus on the ability to capture nuanced information.

Regarding to evaluation of natural language descriptions, low-level speech perception dimensions, either in numerical scores or specific descriptions, are first extracted from the natural language assessments using GPT. Subsequently, the corresponding evaluation metrics are applied to assess each dimension, as shown in Figure \ref{fig5}.

\section{Experimental Results}
\label{results}

\subsection{Open-source auditory LLMs on QualiSpeech Benchmark}

The performance of open-source auditory LLMs on the QualiSpeech benchmark is presented in Table \ref{t1}. Five auditory LLMs, including SALMONN-7B, SALMONN-13B \cite{salmonn}, Qwen-Audio-Chat \cite{qwenaudio}, Qwen2-Audio-7B-Instruct \cite{qwen2audio} and WavLLM \cite{wavllm}, are evaluated. The results show that current open-source auditory LLMs struggle to effectively evaluate speech quality. SALMONN-7B, in particular, exhibits a strong numerical bias, predicting all listening effort and continuity scores as 4, and all noise scores as 3. This tendency towards number preference is also observed, albeit to a lesser extent, in other models. For instance, Qwen-Audio-Chat predicts 80\% of overall quality scores as 3, while Qwen2-Audio-7B-Instruct assigns 66\% of its scores as 3, and WavLLM predicts 86\% of its scores as 4. 

\begin{table*}[t]
\setlength{\tabcolsep}{3pt}
\centering
\begin{tabular}{c|ccccccc}
\toprule
\multirow{2}{*}{\makecell{Training \\ Strategy}} & \multicolumn{7}{c}{Aspect of low-level speech perception} \\
& Noise & Distortion & Speed & Continuity & Effort & Naturalness & Overall \\
\midrule
basic & \textbf{0.721} & 0.553 & \textbf{0.335} & 0.478 & 0.525 & 0.541 & 0.597 \\
balance & 0.696 & 0.547 & 0.268 & 0.458 & 0.497 & 0.540 & 0.600 \\
joint & 0.693 & 0.595 & 0.240 & 0.525 & \textbf{0.578} & 0.565 & 0.636\\
joint + balance & 0.696 & \textbf{0.614} & 0.322 & \textbf{0.535} & \textbf{0.578} & \textbf{0.615} & \textbf{0.660}\\
\bottomrule
\end{tabular}

\vspace{0.15cm}
(a) aspects annotated in scores
\vspace{0.15cm}

\begin{tabular}{c|cccc|cccc|ccc|cc}
\toprule
\multirow{3}{*}{\makecell{Training \\ Strategy}} & \multicolumn{13}{c}{Aspect of low-level speech perception} \\
& \multicolumn{4}{c}{Noise} & \multicolumn{4}{c}{Distortion} & \multicolumn{3}{c}{Unnatural pause} & \multicolumn{2}{c}{Voice} \\
& Prec & Rec & Corr & IoU & Prec & Rec & Corr & IoU  & Prec & Rec & IoU & Corr & GenderAcc \\
\midrule
basic & \textbf{0.70} & 0.50 & \textbf{0.60} & \textbf{0.85} & 0.64 & \textbf{0.97} & \textbf{0.71} & 0.78 & 0.55 & 0.78 & 0.39 & \textbf{0.50} &  0.98\\
balance & 0.66 & 0.53 & 0.57 & 0.82 & \textbf{0.77} & 0.86 & \textbf{0.71} & 0.78 & \textbf{0.57} & 0.78 & 0.37 & 0.49 & 0.98\\
joint & 0.45 & 0.79 & 0.57 & 0.76 & 0.68 & 0.96 & 0.70 & 0.78 & 0.52 & \textbf{0.83} & 0.41 & \textbf{0.50} & 0.98\\
joint + balance & 0.40 & \textbf{0.83} & 0.55 & 0.75 & 0.64 & \textbf{0.97} & 0.70 & \textbf{0.79} & 0.51 & 0.82 & \textbf{0.42} & 0.49 & 0.98 \\
\bottomrule
\end{tabular}

\vspace{0.15cm}
(b) aspects annotated in descriptions
\vspace{-0.1cm}
\caption{Results of learning low-level speech features. PCC is reported for aspects annotated in scores. Correlation scores generated by GPT and IoU of the predicted time period and ground truth are reported for aspects annotated in descriptions. ``Effort'' denotes listening effort and ``GenderAcc'' denotes the accuracy of the classification of the biological gender of the speakers.}
\label{t2}
\vspace{-0.08cm}
\end{table*}

\subsection{Building a speech quality assessment model using QualiSpeech}

We utilize QualiSpeech to build a speech quality assessment auditory LLM, finetuned on SALMONN-7B \cite{salmonn}, with Whisper \cite{whisper} and BEATs \cite{beats} serving as encoder and Vicuna \cite{vicuna} as LLM backbone. We follow the default training configuration in which only the connector speech Q-former and LoRA on LLM are finetuned, as illustrated in Figure \ref{fig5}. First, we fine-tune SALMONN-7B on multiple-choice questions designed to select scores and provide specific descriptions focusing on individual aspects of speech, to assess whether auditory LLMs can effectively understand low-level speech features. In the subsequent phase, we enable the auditory LLMs to generate detailed and logical natural language assessments of speech quality.

\subsubsection{Learning low-level speech features}

Results of learning low-level speech features are shown in Table \ref{t2}. For training strategy, ``basic'' refers to finetuning on a specific task alone, and ``balance'' indicates finetuning on the specific task with balanced distributed data. In this case, the proportions of each score or category (\textit{e.g.}, noise presentation for specific descriptions of noise, and gender for voice) are kept consistent across the dataset. The term ``joint'' denotes joint training across all 11 low-level speech understanding tasks. All models are trained for 10 epochs.

The results show that finetuned auditory LLMs can perform better than the vanilla SALMONN-7B, suggesting that auditory LLMs can, to some extent, grasp low-level speech features. The PCC of noise is around 0.7, indicating the model's ability to distinguish non-semantic components from semantic ones. When comparing the ``basic'' and ``balanced'' training settings, the different data distributions show minimal impact on performance. Joint training improves the classification of certain aspects, such as continuity, without causing significant degradation, suggesting that learning multiple speech features simultaneously does not introduce conflicts. While finetuned auditory LLMs perform reasonably well at predicting speed and noise level, there is still considerable room for improvement in understanding low-level speech features.

\begin{table*}[t]
\setlength{\tabcolsep}{3pt}
\centering
\begin{tabular}{c|ccccccc}
\toprule
\multirow{2}{*}{\makecell{Comments}} & \multicolumn{7}{c}{Aspect of low-level speech perception} \\
& Noise & Distortion & Speed & Continuity & Effort & Naturalness & Overall \\
\midrule
revised concise & 0.656 & \textbf{0.579} & 0.212 & 0.452 & 0.496 & \textbf{0.568} & \textbf{0.630} \\
concise with num & \textbf{0.703} & 0.571 & 0.178 & 0.450 & \textbf{0.513} & 0.535 & 0.622 \\
concise & 0.642 & 0.559 & \textbf{0.263} & \textbf{0.483} & 0.511 & 0.520 & 0.582 \\
detailed & 0.686 & 0.518 & 0.250 & 0.459 & 0.475 & 0.486 & 0.572\\
\bottomrule
\end{tabular}

\vspace{0.15cm}
(a) aspects annotated in scores
\vspace{0.15cm}

\begin{tabular}{c|cccc|cccc|ccc|cc}
\toprule
\multirow{3}{*}{\makecell{Comments}} & \multicolumn{13}{c}{Aspect of low-level speech perception} \\
& \multicolumn{4}{c}{Noise} & \multicolumn{4}{c}{Distortion} & \multicolumn{3}{c}{Unnatural pause} & \multicolumn{2}{c}{Voice} \\
& Prec & Rec & Corr & IoU & Prec & Rec & Corr & IoU  & Prec & Rec & IoU & Corr & GenderAcc \\
\midrule
revised concise & 0.40 & 0.50 & 0.49 & 0.49 & \textbf{0.80} & 0.74 & 0.67 & \textbf{0.78} & 0.54 & 0.52 & \textbf{0.34} & 0.48 & 0.97\\
concise with num & \textbf{0.62} & \textbf{0.54} & \textbf{0.53} & \textbf{0.73} & 0.76 & \textbf{0.84} & 0.66 & 0.77 & 0.55 & \textbf{0.60} & \textbf{0.34} & 0.48 & \textbf{0.98}\\
concise & 0.34 & 0.52 & 0.49 & 0.41 & 0.74 & 0.81 & 0.65 & 0.73 & 0.58 & 0.57 & 0.30 & 0.49 & 0.96\\
detailed & 0.50 & 0.27 & 0.46 & 0.61 & 0.76 & 0.74 & \textbf{0.68} & 0.73 & \textbf{0.60} & 0.56 & 0.33 & \textbf{0.51} & \textbf{0.98}\\
\bottomrule
\end{tabular}

\vspace{0.15cm}
(b) aspects annotated in descriptions
\vspace{-0.1cm}
\caption{Results of learning natural language descriptions. PCC is reported for aspects annotated in scores. Correlation scores generated by GPT and IoU of predicted time period and ground truth are reported for aspects annotated in descriptions. ``Effort'' denotes listening effort and ``GenderAcc'' denotes the accuracy of the classification of the biological gender of the speakers.}
\label{t4}
\end{table*}

\begin{table}[t]
\setlength{\tabcolsep}{3pt}
\centering
\begin{tabular}{c|cc}
\toprule
Model & Vicuna-v1.5-7B & GPT-4o-mini \\
\midrule
Acc & 0.28 & 0.46 \\
\bottomrule
\end{tabular}

\caption{Results on the reasoning for the overall quality score of text LLMs based on basic groundtruth low-level speech features.}
\label{t5}
\vspace{-0.08cm}
\end{table}

Results for specific descriptions, which are unique to natural language assessment, are more promising. An IoU score of approximately 0.8 is achieved when describing noise or distortion, indicating that the model always identifies the correct time periods if the model recognizes noise or distortion. This underscores the potential of using auditory LLMs to generate nuanced and detailed descriptions of speech quality. The correlation scores are also satisfactory, demonstrating that the model can accurately describe the type of noise or distortion in most cases. However, the precision and recall metrics reveal that the model's ability to reliably detect the presence of noise or distortion still requires improvement. A simple balancing of data distribution, in this case, does not yield significant benefits. When it comes to identifying unnatural pauses, the results are less favourable compared to noise and distortion, suggesting rhythm is hard to learn. For voice description, while the model performs excellently in gender classification, it struggles with age and tone, resulting in a moderate correlation score of around 0.5.

We further assess the generalization ability of auditory LLMs across different data types, with the results presented in Table \ref{th} in Appendix \ref{append:h2}. In this analysis, we finetune the models on one specific data type and evaluate their performance on all data types. The results show that if a model is trained only on one data type, it will exhibit a poor generalization to other domains. The findings also show that incorporating more diverse data sources leads to improved performance across all domains. The results suggest that building a robust and general speech quality assessment model for both synthetic and real data is feasible, highlighting the importance of datasets like QualiSpeech, which encompass a wide array of data sources.

\subsubsection{Learning natural language descriptions}

We also investigate whether auditory LLMs can benefit from reasoning in natural language. So we further finetune the checkpoint jointly trained on all basic low-level understanding tasks to generate descriptive comments, This process involves analyzing each aspect individually before synthesizing all the dimensions to derive an overall score. All models were trained for 10 epochs. All models are trained for 10 epochs. Different formats are also explored, with detailed settings and examples provided in the Appendix \ref{append:d}.

Results shown in Table \ref{t4} show that auditory LLMs can generate a paragraph of natural language speech quality assessment, achieving accuracy comparable to evaluating each aspect separately. The length of the generated text has minimal impact on performance. Including numerical scores within the natural language description enhances the quality of the output, likely due to the added specificity of the information provided. Revising the generated comments to eliminate any hallucinations is crucial for producing high-quality, reliable speech quality assessments.

However, the model does not achieve higher accuracy in overall score prediction through reasoning in natural language. Incorrectly predicted low-level speech aspects also interference with reasoning. To address this, we experiment with text-based LLM reasoning for predicting the overall quality score, using the groundtruth low-level speech features. The results, shown in Table \ref{t5}, reveal that the LLM backbone of SALMONN-7B, Vicuna-v1.5-7B, lags behind all fine-tuned auditory LLMs, suggesting that the failure of reasoning is partly due to the LLM backbone's weak reasoning capabilities. In contrast, GPT-4o-mini outperforms all fine-tuned auditory LLMs, highlighting the possibility of reasoning when assessing speech quality. 

\section{Conclusion}

In this paper, we present QualiSpeech, a comprehensive speech quality dataset curated from diverse sources, encompassing a wide range of aspects and incorporating natural language descriptions. We also introduce the QualiSpeech Benchmark, designed to evaluate the low-level speech understanding capabilities of auditory LLMs. Benchmark results reveal that current open-source auditory LLMs face challenges in accurately assessing speech quality. Our experiments show that natural language descriptions provide more detailed insights into noise and distortion compared to traditional methods. Generalization experiments highlight the importance of incorporating data from diverse sources to develop a robust, general-purpose speech quality assessment model suitable for all scenarios. We hope QualiSpeech will inspire further research into natural language-based speech quality assessment, enabling more fine-grained and reliable evaluations.


\section*{Acknowledgments}

This work was supported by a research grant from ByteDance in 2024-2025, and Natural Science Foundation of China (62476151).

\section*{Limitations}

There are some limitations of our work. First, although QualiSpeech encompasses multiple aspects and diverse sources, it inevitably leaves some aspects and sources uncovered. Second, each speech sample in QualiSpeech is annotated by only one listener due to the complexity of the annotation process, whereas MOS scoring typically involves evaluations from multiple listeners per sample. Despite this limitation, we believe our current dataset provides a valuable foundation for the community to explore and develop initial approaches. Lastly, our fine-tuned auditory LLMs do not yet fully leverage reasoning in natural language, primarily due to the limitations of the underlying LLM backbone. We hope future stronger auditory LLMs can utilize and benefit from reasoning in natural language descriptions of QualiSpeech.

\section*{Ethics Statement}

All the models and datasets in this paperare publicly accessible and used under licenses. As for the generated speech samples from recent TTS systems in QualiSpeech, all open-source TTS models are used under corresponding licenses. Our dataset is provided under a Creative Commons Attribution-NonCommercial-ShareAlike 4.0 International License. All annotators received clear annotation rules before the annotation and fair payments upon completion. We believe the auditory LLM evaluator can only be used as a reference and should not replace human evaluators.
\bibliography{custom}

\clearpage
\newpage
\appendix

\section{Speech quality assessment: MOS and natural language}
\label{append:a}

We believe that describing speech quality in natural language and evaluating it using numerical scores, such as MOS, are complementary approaches rather than direct alternatives. Therefore, natural language assessment is not compared with traditional numerical score method in our paper. MOS score is a rough overall assessment that facilitates easy comparisons between samples and systems, while natural language provides more detailed and specific information, such as identifying the type and occurrence time of distortions. If the goal is to simply compare the quality of two speech samples, MOS remains the preferred choice. However, for more detailed and instructive feedback — such as identifying consistent issues like a clicking sound at the beginning of TTS outputs — natural language assessment is invaluable. Additionally, audio LLMs can reason in natural language, providing explanations for overall quality judgments, which paves the way for more reliable speech quality assessments.

\section{Choice of low-level speech aspects}
\label{append:j}

We aim to build a comprehensive low-level speech quality assessment dataset, so we refer to different aspects discussed in previous works, for example, noise\cite{nisqa,choice2}, distortion\cite{nisqa,choice2}, speed\cite{choice3}, naturalness\cite{choice4,scale}, listening effort\cite{scale,choice6}, continuity\cite{nisqa}, with the hope that low-level speech aspects in QualiSpeech can cover not only basic degradations, such as background noise and distortion, but also subjective perceptual assessment, including speech speed, continuity, naturalness and listening effort. We also collect specific descriptions to further enrich the description perspectives, building a speech quality assessment dataset containing comprehensive low-level speech features and natural language descriptions.

\section{Human annotators information}
\label{append:c}

Our human annotators consist of 21 females and 4 males, all in their 20s. Due to local constraints, all annotators are native Mandarin speakers with English proficiency equivalent to an IELTS score of 7.0 or higher, although they are not native English speakers. And strong correlations between high level non-native listeners and native English listeners are reported in \cite{vcc}. Each annotation takes approximately six minutes to complete, and annotators are paid 6 Chinese Yuan (approximately 0.8 USD) per sample, satisfying the minimum income standards of our region.

\section{Label distributions of QualiSpeech}
\label{append:i}

\begin{figure*}[h]
\centering
\includegraphics[width=\textwidth]{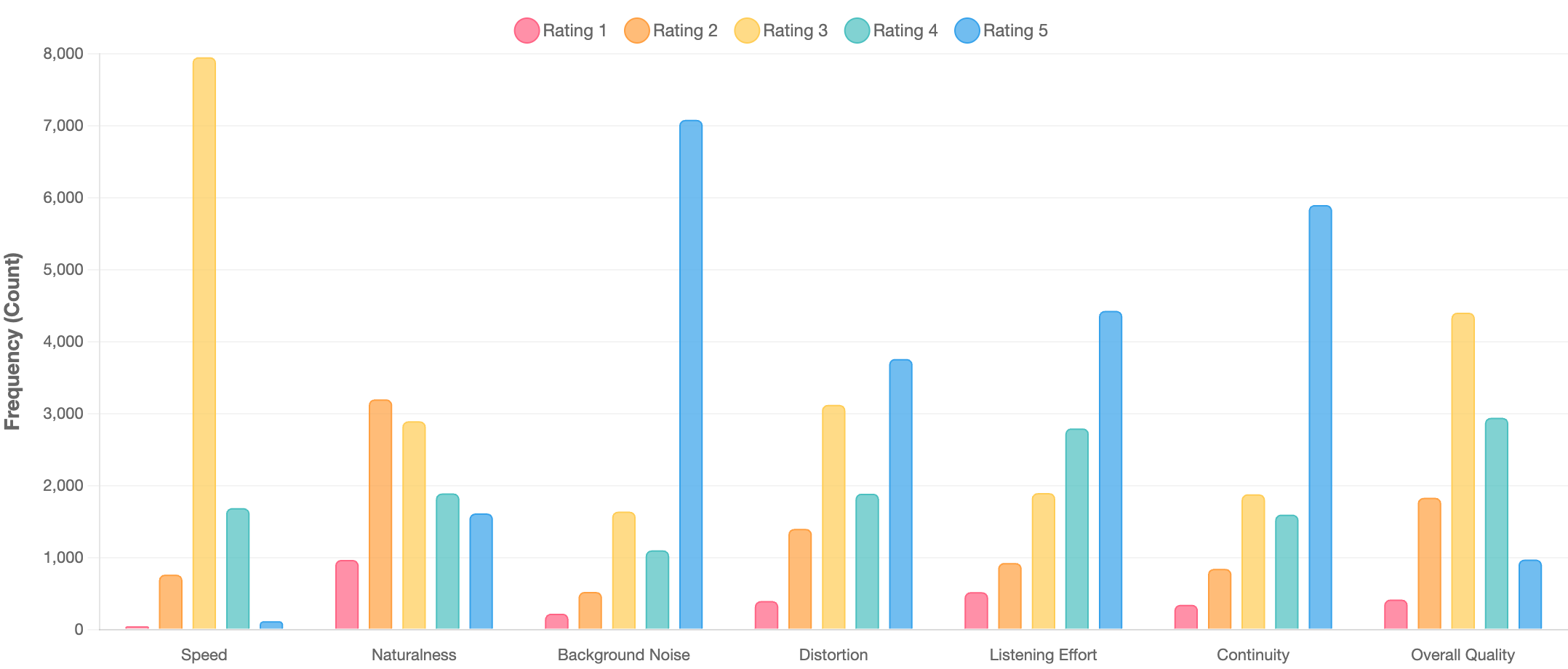}
\caption{Label distributions of aspects annotated in scores in QualiSpeech}
\label{fig_dis1}
\end{figure*}

The label distributions of aspects annotated in scores are shown in Figure \ref{fig_dis1}. The distributions of speed and overall quality roughly follow a normal pattern. For noise, continuity, and distortion, part of the samples are noisy or discontinuous, forming an approximately normal distribution, while the other half are clean or smooth rated as 5. Listening effort exhibits a ladder-like distribution, with the highest number of samples receiving a score of 5 and the lowest number receiving a score of 1. naturalness follows a shifted normal distribution, peaking at 2 rather than 3.

\section{Detailed statistics of QualiSpeech}
\label{append:b}

The detailed statistics of QualiSpeech dataset shown in Table \ref{tb}. 

\begin{table}[h]
\vspace{0.1cm}
\setlength{\tabcolsep}{3pt}
\centering
\begin{tabular}{cc|p{1.0cm}<{\centering}p{1.0cm}<{\centering}p{1.0cm}<{\centering}}
\toprule
\multirow{2}{*}{Type} & \multirow{2}{*}{Source} & \multicolumn{3}{c}{\# Utterances} \\
 & & Train & Valid & Test \\ 
\midrule
\multirow{2}{*}{\makecell{Synthetic}} & BVCC & 4546 & 975 & 912  \\
& Recent TTS & 600 & 60 & 60\\
\midrule
\multirow{4}{*}{\makecell{Real}} & GigaSpeech & 2000 & 200 & 400 \\
& NISQA LIVE & 529 & 106 & 0\\
& NISQA FOR & 0 & 0 & 180\\
& NISQA P501 & 0 & 0 & 180\\
\midrule
\multirow{3}{*}{\makecell{Simulated \\ Real}} & NISQA SIM & 2883 & 826 & 0\\
& NISQA FOR & 0 & 0 & 60 \\
& NISQA P501 & 0 & 0 & 60 \\
\midrule
\multicolumn{2}{c|}{Summary of synthetic} & 5146 & 1035 & 972 \\
\multicolumn{2}{c|}{Summary of all real} & 5412 & 1132 & 880 \\
\multicolumn{2}{c|}{Summary} & 10558 & 2167 & 1852 \\

\bottomrule
\end{tabular}
\caption{Statistics of our QualiSpeech dataset, with respect to data sources and dataset splits.}
\label{tb}
\end{table}

The Pearson correlation coefficient (PCC) between overall quality scores in QualiSpeech and MOS scores in the source datasets is calculated to check the consistency of QualiSpeech with the source datasets, as shown in Table \ref{tb2}. It is quite relevant between these two scores, considering that overall quality scores in QualiSpeech are integers while MOS scores in the source datasets are decimals.

\begin{table}[h]
\vspace{0.1cm}
\setlength{\tabcolsep}{3pt}
\centering
\begin{tabular}{c|cc}
\toprule
Subset & BVCC test split & NISQA TEST \\
\midrule
PCC & 0.55 & 0.66 \\
\bottomrule
\end{tabular}
\caption{The correlation scores between overall quality scores in QualiSpeech and MOS scores in the source dataset.}
\label{tb2}
\end{table}

\section{Experiment setup for descriptive comments}
\label{append:d}

``Concise'' and ``detailed'' denotes the text length of the descriptive comments. For ``concise'', a maximum limitation of 150 words is added for generated concise descriptions, while for detailed descriptions there is no maximum limitation of text length. For concise descriptions, the average text length is 136. While for detailed descriptions, it is 209. ``Concise with num'' means the annotated scores will be included in the generated descriptions and the generation has a maximum limitation of 150 words. ``Revised concise'' is the revised version of concise descriptions by human annotators (annotation step 3), for other descriptions, they are not revised. Below is a example of different kinds of descriptions:

\textbf{Concise}:
\textit{The speech sample presents a pleasant listening experience with background noise being virtually non-existent, allowing for clear audio. However, there is some noticeable pitch distortion accompanied by a mechanical sound that detracts from the overall quality, making the speech somewhat unnatural. Despite this distortion, the audio remains smooth and free of interruptions, contributing positively to the listening experience. The listener requires only a moderate amount of effort to comprehend the content, indicating that the speech is mostly accessible. The speaking speed is moderate, which complements the gentle and peaceful tone of the young man's voice, enhancing the overall impression. While the naturalness of the speech is lacking, the combination of smooth delivery and minimal background noise results in a fair overall quality, suggesting room for improvement in the clarity and authenticity of the voice.}

\textbf{Revised concise}:
\textit{The speech sample presents a pleasant listening experience with background noise being virtually non-existent, allowing for clear audio. However, there is some noticeable pitch distortion accompanied by a mechanical sound that detracts from the overall quality from 0 to 2 seconds, making the speech somewhat unnatural. Despite this distortion, the audio remains smooth and free of interruptions, contributing positively to the listening experience. The listener requires only a moderate amount of effort to comprehend the content, indicating that the speech is mostly accessible. The speaking speed is moderate, which complements the gentle and peaceful tone of the young man's voice, enhancing the overall impression. While the naturalness of the speech is lacking, the combination of smooth delivery and minimal background noise results in a fair overall quality, suggesting room for improvement in the clarity and authenticity of the voice.}

\textbf{Concise with num}:
\textit{The speech sample exhibits a background noise level rated at 5, indicating that it is not noticeable, which positively contributes to the overall quality. However, distortion is rated at 3, suggesting some pitch distortion and mechanical sound present throughout the voice, which detracts from the naturalness of the speech, rated at 2, indicating it sounds fairly unnatural. Discontinuity is rated at 5, reflecting a very smooth delivery, while listening effort is rated at 4, meaning only attention is necessary to understand the content. The speaking speed is moderate at 3, which is acceptable. The voice is described as a young man's gentle voice with a peaceful tone, adding a pleasant feeling. Overall, the speech quality is rated at 3, indicating a fair quality, primarily impacted by the distortion and naturalness issues, despite the smoothness and low background noise.}

\textbf{Detailed}:
\textit{The speech sample presents a range of qualities that contribute to its overall evaluation. Starting with background noise, it is commendable that there is no noticeable interference, allowing the listener to focus on the content without distraction. However, the distortion aspect reveals some concerns, as there is a pitch distortion accompanied by a mechanical sound that persists throughout the initial two seconds, which detracts from the natural quality of the speech. Despite this, the discontinuity is excellent, with the audio being very smooth, indicating that there are no breaks or stutters that could disrupt comprehension. In terms of listening effort, the sample requires some attention but does not demand significant effort, suggesting that the listener can grasp the meaning with relative ease. The naturalness of the speech, however, is somewhat lacking, as it does not closely resemble natural human speech, which may affect the listener's engagement. The speaking speed is moderate, which is appropriate for comprehension, neither too fast nor too slow. The voice itself is described as a young man's gentle voice with a peaceful tone, which adds a positive emotional layer to the experience. Overall, while the speech has strengths in background noise management, smoothness, and listener engagement, the distortion and naturalness issues bring down its quality. Therefore, the overall quality can be considered fair, reflecting a mix of commendable attributes and notable shortcomings.}

\section{Text instructions for Parler-TTS}
\label{append:e}
The text instructions for Parler-TTS which specfies speaker are sampled from a corpus containing 44 sentences. 10 sentences are generated by GPT following the format of the example instruction\textsuperscript{\ref{parlertts}}, with age, gender and speed changed. 34 sentences are simple ``someone's voice'' in which \textit{someone} refers to an available built-in speaker. The full list of text instructions are presented as below:

1. \textit{A young male speaker presents a calm, steady speech at a slightly slower-than-average speed, with a deeper pitch. The recording quality is excellent, with his voice sounding close up and clear, as if he’s speaking directly into the microphone.}

2. \textit{An elderly female speaker delivers a warm, gentle speech, characterized by a slow pace and soft pitch. Her voice is clear and intimate, with the high-quality recording capturing her subtle inflections.}

3. \textit{A middle-aged male speaker gives a confident and slightly assertive speech, maintaining a steady pace with a moderate pitch. The recording is pristine, and his voice sounds rich and detailed, as if he’s speaking in a small, quiet room.}

4. \textit{A young female speaker offers a friendly and enthusiastic speech with a faster-than-average speed and a higher pitch. Her voice is clear and lively, captured in great quality, creating a vibrant and engaging listening experience.}

5. \textit{A middle-aged female speaker delivers a formal, measured speech at a moderate speed, with a soft but clear voice. The recording quality is high, capturing her voice with warmth and precision, as if in a professional studio.}

6. \textit{A male teenager speaks in a casual, conversational tone, with a moderate speed and a slightly higher pitch than average. The recording is of very high quality, with his voice sounding crisp and close, capturing even his smallest breaths.}

7. \textit{An elderly male speaker gives a slow, reflective speech with a slightly lower pitch, exuding wisdom and calm. The recording quality is excellent, making his voice sound intimate and detailed, as if he’s speaking directly to the listener.}

8. \textit{A young adult female speaks with high energy, delivering her message at a rapid pace and with a high pitch. The quality of the recording is very clear, emphasizing her upbeat tone and the subtleties in her speech.}

9. \textit{A young adult male delivers a slightly monotone, yet articulate, speech with a moderate speed and pitch. The recording quality is excellent, making his voice sound clear and close, as if he’s speaking in a quiet room.}

10. \textit{A middle-aged female speaker speaks with a gentle, soothing tone, maintaining a slow pace and lower pitch. Her voice is captured in exceptional quality, with every word resonating warmly and clearly, giving an intimate listening experience.}

11. \textit{Laura's Voice.}

12. \textit{Gary's Voice.} 

13. \textit{Jon's Voice.} 

14. \textit{Lea's Voice.} 

15. \textit{Karen's Voice.} 

16. \textit{Rick's Voice.} 

17. \textit{Brenda's Voice.} 

18. \textit{David's Voice.} 

19. \textit{Eileen's Voice.} 

20. \textit{Jordan's Voice.} 

21. \textit{Mike's Voice.} 

22. \textit{Yann's Voice.} 

23. \textit{Joy's Voice.} 

24. \textit{James's Voice.} 

25. \textit{Eric's Voice.} 

26. \textit{Lauren's Voice.} 

27. \textit{Rose's Voice.} 

28. \textit{Will's Voice.} 

29. \textit{Jason's Voice.} 

30. \textit{Aaron's Voice.} 

31. \textit{Naomie's Voice.} 

32. \textit{Alisa's Voice.} 

33. \textit{Patrick's Voice.} 

34. \textit{Jerry's Voice.} 

35. \textit{Tina's Voice.} 

36. \textit{Jenna's Voice.} 

37. \textit{Bill's Voice.} 

38. \textit{Tom's Voice.} 

39. \textit{Carol's Voice.} 

40. \textit{Barbara's Voice.} 

41. \textit{Rebecca's Voice.} 

42. \textit{Anna's Voice.} 

43. \textit{Bruce's Voice.} 

44. \textit{Will's Voice.}

\clearpage
\section{Examples in QualiSpeech}
\label{append:i}

Example 1 (a low quality synthetic speech sample mixed with baby crying):

{\itshape Speed: 3

Naturalness: 2

Background noise: 3	

Distortion: 4	

Listening effort: 3	

Continuity: 2	

Overall quality: 2

Feeling of voice: the calm and robotically mechnical voice of a young woman

Noise description: There is outdoor background noise like baby crying in the audio from 0 to 3.3 seconds.

Distortion description: There is a pitch distortion with mechnical sound in 2.5-3s.

Unnatural pause: 2.2-2.5s. 

Descriptive comments: The speech sample presents a moderate level of background noise, with outdoor sounds, including a baby crying, noticeable from 0 to 3.3 seconds, which somewhat distracts from the clarity of the speech. There is slight distortion, particularly a mechanical pitch alteration occurring between 2.5 and 3 seconds, which affects the naturalness of the voice. The audio experiences some discontinuity, making it fairly disjointed, especially around 2.2 to 2.5 seconds, where the unnatural quality is evident. Listeners may find moderate effort required to comprehend the speech due to these factors. The speaking speed is moderate, contributing to a calm delivery, but the overall quality remains poor due to the combination of these issues. The voice, described as that of a middle-aged man, has a calm yet somewhat mechanical tone, which further detracts from the overall listening experience.}

\newpage
Example 2 (a high quality audiobook data):

{\itshape Speed: 3

Naturalness: 5

Background noise: 5	

Distortion: 5	

Listening effort: 5	

Continuity: 5	

Overall quality: 5

Feeling of voice: A middle-aged man with bright voice that makes people feel relaxed

Noise description: not noticeable.

Distortion description: not distorted.

Unnatural pause: very smooth.

Descriptive comments: The speech sample exhibits exceptional quality across all evaluated aspects. The background noise is virtually non-existent, allowing for a clear listening experience. There are no distortions present, ensuring that the natural sound of the speech is preserved. The audio flows very smoothly, with no interruptions or stutters, contributing to a seamless listening experience. Understanding the content requires no effort, as the speech is delivered in a way that is easily comprehensible. The naturalness of the voice is striking, closely resembling human speech in tone and clarity. The speaking speed is moderate, which balances well with the overall delivery. The voice of the speaker, described as that of a middle-aged man with a bright tone, evokes a sense of relaxation in listeners. Overall, this speech sample is of excellent quality, making it a pleasure to listen to.}

\section{Questionnaires and prompts}
\label{append:f}

\begin{figure*}[h]
\centering
\includegraphics[width=0.7\textwidth]{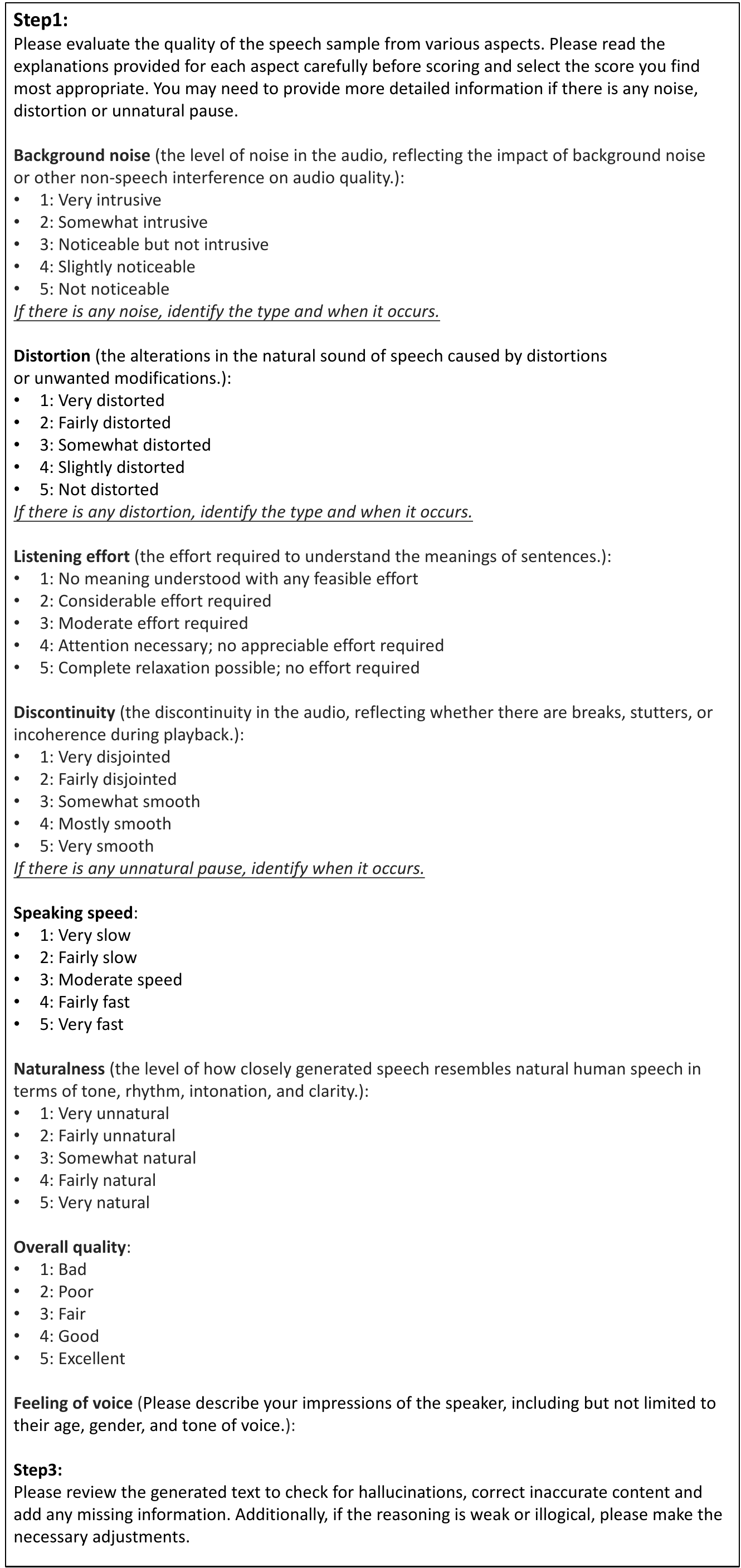}
\caption{Questionnaires and instructions for annotation}
\label{fig10}
\end{figure*}

\begin{figure*}[h]
\centering
\includegraphics[width=\textwidth]{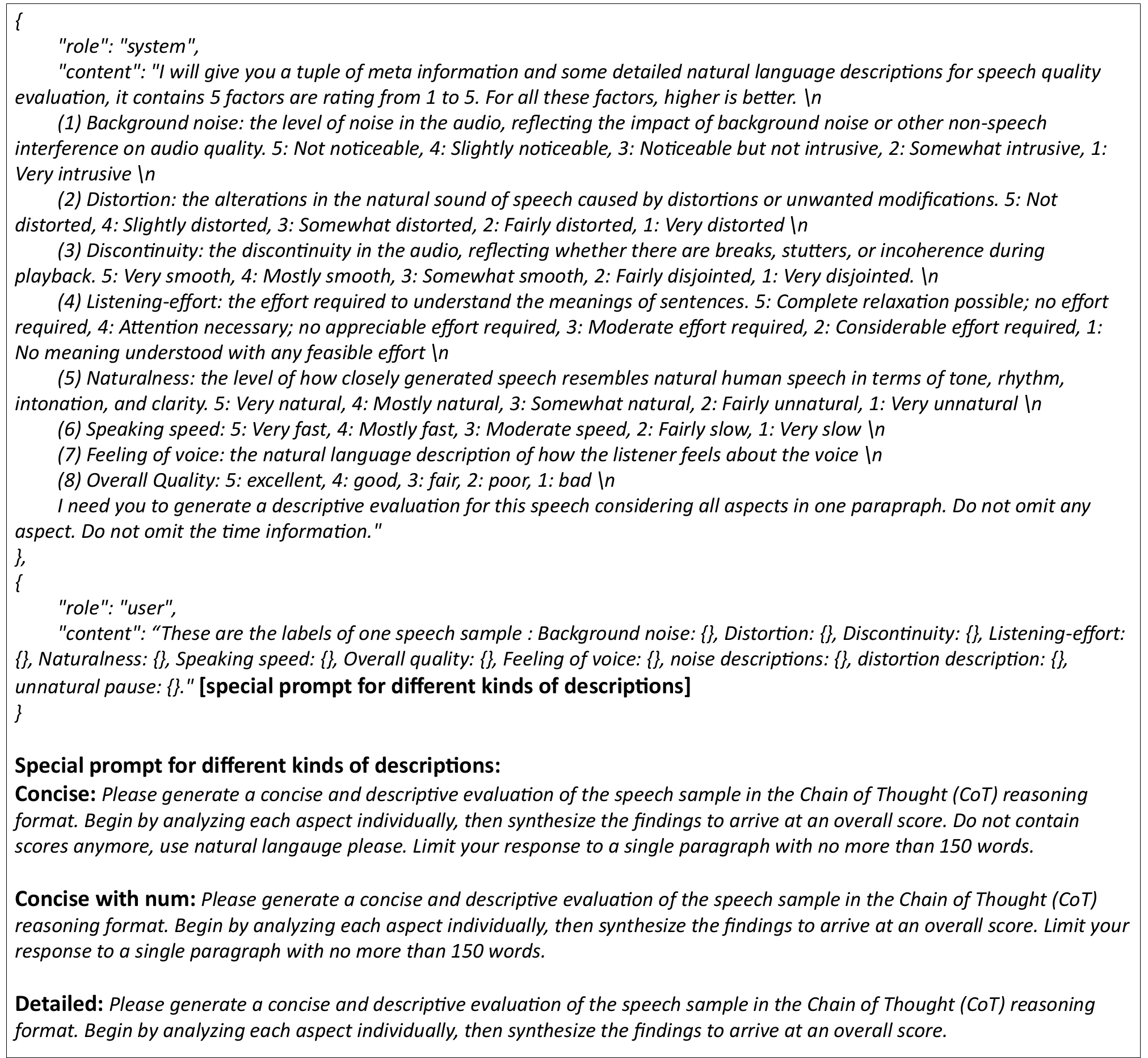}
\caption{Prompts used in annotation procedure step 2 for generating different kinds of descriptive comments}
\label{fig6}
\end{figure*}

\begin{figure*}[h]
\centering
\includegraphics[width=\textwidth]{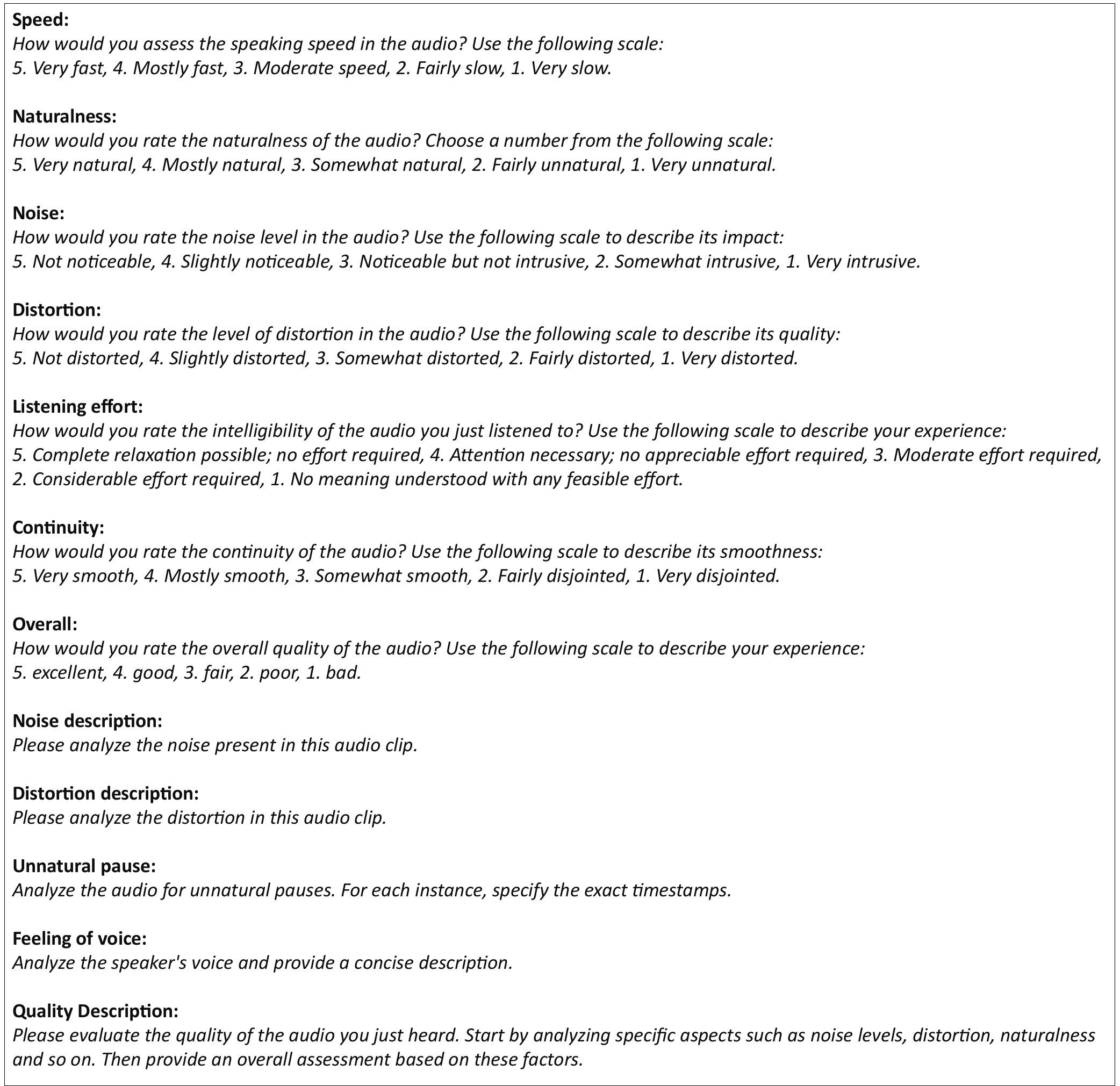}
\caption{Prompts for each tasks when finetuning and testing auditory LLMs.}
\label{fig9}
\end{figure*}

\begin{figure*}[h]
\centering
\includegraphics[width=\textwidth]{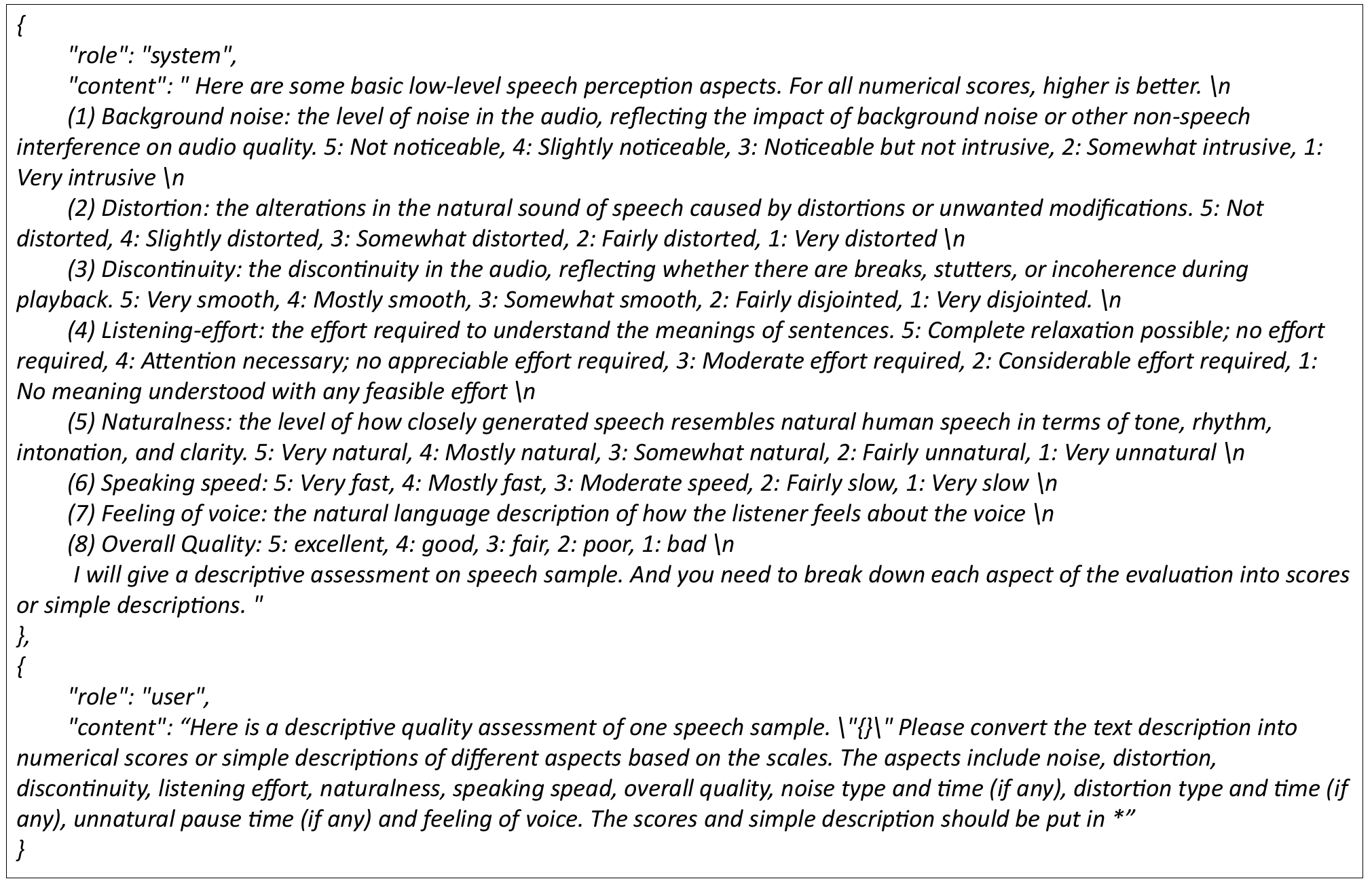}
\caption{Prompts for breaking down description comments into basic aspects, used in evaluating natural language descriptions.}
\label{fig7}
\end{figure*}

\begin{figure*}[h]
\centering
\includegraphics[width=\textwidth]{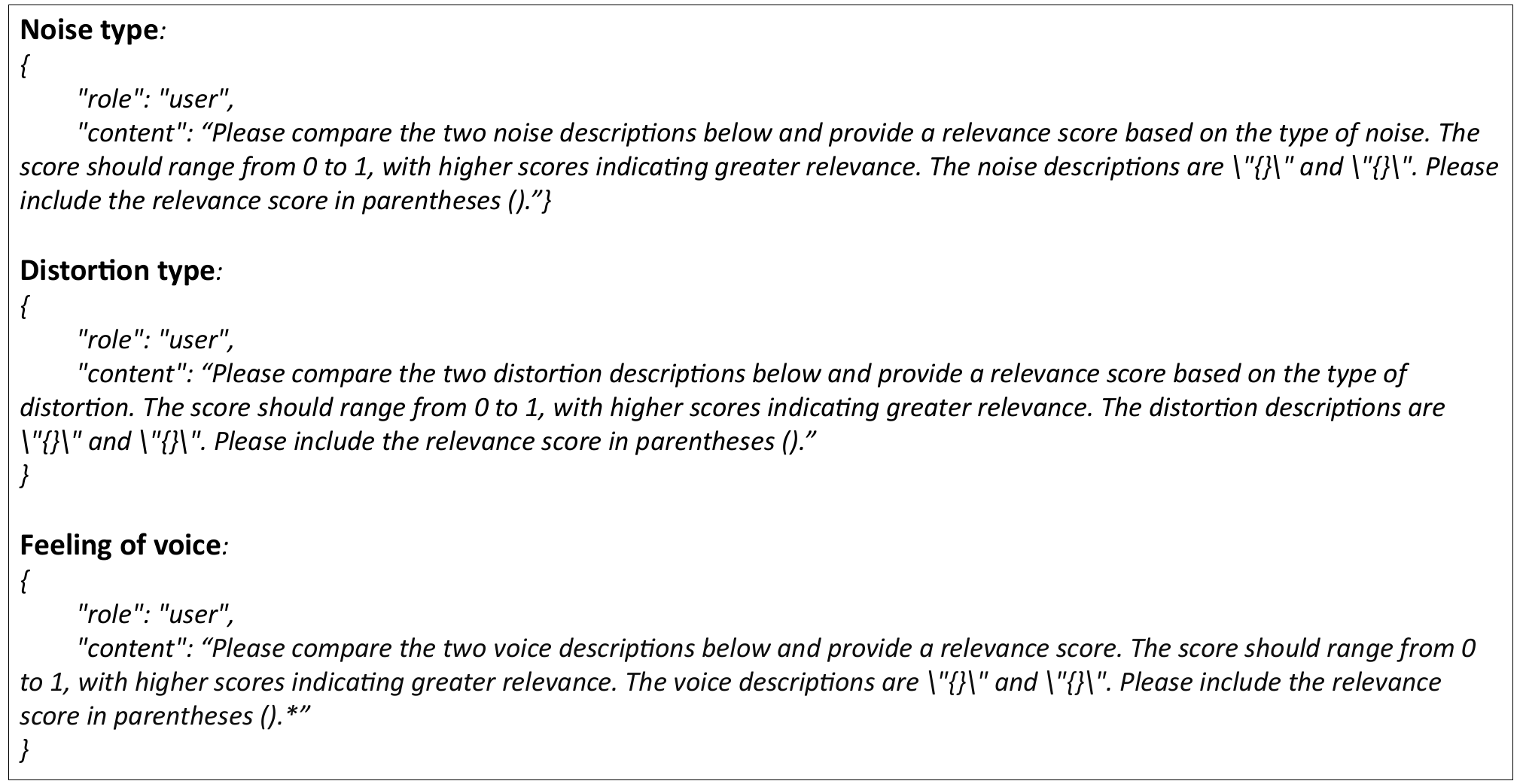}
\caption{Prompts for calculating correlation score between two descriptions, used in evaluating specific descriptions.}
\label{fig8}
\end{figure*}

\clearpage
\section{More results}
\label{append:h}

\subsection{Ablation study on finetuning encoders}

We also investigate the influence of finetuning encoders, with results demonstrated in Table \ref{th2}. The training setting follows the ``joint'' setting. Results show that finetuning encoders can bring further performance improvement. We suggest that with encoders joint finetuned, the low-level speech information can be transformed to a format that LLM backbone can better understand, resulting in a performance gain.

\subsection{Results of generalization experiments}
\label{append:h2}

\begin{table*}[b]
\setlength{\tabcolsep}{3pt}
\centering
\begin{tabular}{c|ccccccc}
\toprule
\multirow{2}{*}{\makecell{Finetuning \\ Encoder}} & \multicolumn{7}{c}{Aspect of low-level speech perception} \\
& Noise & Distortion & Speed & Continuity & Effort & Naturalness & Overall \\
\midrule
freeze & 0.693 & 0.595 & 0.240 & 0.525 & 0.578 & 0.565 & 0.636\\
Whisper  & 0.687 & 0.626 & \textbf{0.355} & 0.576 & 0.598 & 0.587 & 0.643 \\
BEATs & \textbf{0.723} & 0.645 & 0.308 & 0.536 & 0.597 & 0.595 & 0.654 \\
Whisper + BEATs & 0.707 & \textbf{0.648} & 0.325 & \textbf{0.601} & \textbf{0.617} & \textbf{0.606} & \textbf{0.670}\\
\bottomrule
\end{tabular}

\caption{Results of ablation study on finetuning encoder}
\label{th2}
\vspace{-0.08cm}
\end{table*}

\begin{table*}[b]
\setlength{\tabcolsep}{3pt}
\centering
\begin{tabular}{c|cccccccc}
\toprule
\multirow{2}{*}{\makecell{Training \\ Dataset}} & \multicolumn{7}{c}{Aspect of low-level speech perception} \\
& Noise & Distortion & Speed & Continuity & LE & Naturalness & Average first 6 & Overall\\
\midrule
synthetic & 0.785 & 0.533 & \textbf{0.288} & 0.324 & 0.517 & 0.426 & \textbf{0.479} & 0.520\\
real & 0.802 & 0.468 & 0.211 & 0.163 & 0.476 & 0.450 & 0.428 & 0.471\\
all & \textbf{0.806} & \textbf{0.543} & 0.200 & \textbf{0.333} & \textbf{0.521} & \textbf{0.463} & 0.478 & \textbf{0.543}\\
\bottomrule
\end{tabular}

\vspace{0.15cm}
(a) Results on synthetic data
\vspace{0.15cm}

\begin{tabular}{c|cccccccc}
\toprule
\multirow{2}{*}{\makecell{Training \\ Dataset}} & \multicolumn{7}{c}{Aspect of low-level speech perception} \\
& Noise & Distortion & Speed & Continuity & LE & Naturalness & Average first 6 & Overall \\
\midrule
synthetic & \textbf{0.717} & 0.458 & 0.221 & 0.523 & 0.500 & 0.399 & 0.470 & 0.597\\
real & 0.652 & 0.588 & \textbf{0.284} & 0.607 & 0.582 & 0.614 & 0.555 & 0.709 \\
all & 0.589 & \textbf{0.611} & 0.258 & \textbf{0.691} & \textbf{0.631} & \textbf{0.618} & \textbf{0.566} & \textbf{0.724}\\
\bottomrule
\end{tabular}

\vspace{0.15cm}
(b) Results on real data
\vspace{-0.1cm}
\caption{Full results of generalization experiments. ``Average first 6'' denotes the averaged accuracy of 6 aspects annotated in scores except for overall.}

\label{th}
\end{table*}

\newpage
\subsection{Holistic correlation of generated descriptions}

We also try to give a holistic correlation score generated by GPT based on the entire description, the results are shown below. The results show that a holistic correlation score makes it hard to distinguish the differences in methods, and therefore correlation scores of specific descriptions are reported in Section \ref{results}.

\begin{table}[h]
\setlength{\tabcolsep}{3pt}
\centering
\begin{tabular}{c|c}
\toprule
Comments & Holistic Corr. \\
\midrule
revised concise & 0.667 \\
concise with num & 0.675 \\
concise & 0.639 \\
detailed & 0.661 \\
\bottomrule
\end{tabular}
\caption{Holistic correlation of the entire generated descriptions}
\label{tj3}
\vspace{-0.08cm}
\end{table}

\newpage
\subsection{Investigation of multiple annotations}

To investigate the impact of multiple annotations, we collect 2 more annotations for a subset of test split in QualiSpeech (only the data from BVCC resource). We further explore the consistency among multiple annotations and assess whether incorporating multiple annotations can benefit the testing procedure.

\subsubsection{Consistency of multiple annotations}

To evaluate the consistency of multiple annotations, mutual consistency metrics are calculated by averaging pairwise consistency metrics. For aspects annotated in scores, PCC is selected as consistency metric. For aspects annotated in descriptions, correlation score generated by GPT is utilized.

Results show that different annotations do not exhibit a high consistency, underscoring the inherent complexity of speech quality assessment. Notably, consistency is higher for basic degradations such as noise and distortion compared to more subjective perceptual aspects. Furthermore, compared results in Table \ref{t2} and Table \ref{tj41}, the finetuned auditory LLM demonstrates a promising potential to function as an annotator for speech quality evaluation.

\begin{table}[h]
\setlength{\tabcolsep}{3pt}
\centering
\begin{tabular}{c|c}
\toprule
Low-level & \multirow{2}{*}{Mutual PCC}  \\
speech aspect & \\
\midrule
noise & 0.728 \\
distortion & 0.682 \\
speed & 0.316 \\
continuity & 0.604 \\
effort & 0.653 \\
naturalness & 0.458 \\
overall & 0.603 \\
\bottomrule
\end{tabular}
\caption{Consistency of multiple annotations on aspects annotated in scores}
\label{tj41}
\vspace{-0.08cm}
\end{table}

\begin{table}[h]
\setlength{\tabcolsep}{3pt}
\centering
\begin{tabular}{c|c}
\toprule
Low-level & \multirow{2}{*}{Mutual Corr}  \\
speech aspect & \\
\midrule
noise & 0.672 \\
distortion & 0.611 \\
voice & 0.483\\
\bottomrule
\end{tabular}
\caption{Consistency of multiple annotations on aspects annotated in descriptions}
\label{tj42}
\vspace{-0.08cm}
\end{table}

\newpage
\subsubsection{Multiple annotations for evaluation}

We further explore the use of multiple annotations for evaluation. The model tested is the one trained under ``joint'' setting in Table \ref{t2}. We report the evaluation metrics for each annotation individually, along with their averaged results. Furthermore, we explore two new evaluation metrics. For aspects annotated in scores, we use the mean of all annotations as the ground truth label, denoted as ``Mean value" in Table \ref{tj43}. For aspects annotated in descriptions, we select the highest correlation score as the final result, labeled as ``Best" in Table \ref{tj44}.

The results indicate that incorporating multiple annotations enables the use of more reliable evaluation metrics. For aspects assessed with numerical scores, the PCC with the mean value is the highest, suggesting that even when the model is trained on a single annotation, it aligns most closely with the averaged scores. This finding highlights that averaging multiple annotations will lead to more stable evaluation. For aspects annotated in descriptions, we believe ``Best'' is also a better evaluation metric, since generating descriptions is an open-ended task with multiple reasonable answers. We hope our pioneering dataset can inspire large-scale natural language speech quality assessment dataset with multiple annotations in the future.

\begin{table}[h]
\setlength{\tabcolsep}{3pt}
\centering
\resizebox{0.48\textwidth}{!}{
\begin{tabular}{c|ccccc}
\toprule
Low-level & \multirow{2}{*}{1} & \multirow{2}{*}{2} & \multirow{2}{*}{3} & \multirow{2}{*}{Average} & Mean \\
speech aspect & & & & & value \\
\midrule
noise & 0.756 & 0.755 & 0.748 & 0.753 & 0.831 \\
distortion & 0.704 & 0.709 & 0.693 & 0.702 & 0.790 \\
speed & 0.183 & 0.310 & 0.346 & 0.280 & 0.393 \\
continuity & 0.642 & 0.635 & 0.608 & 0.628 & 0.731 \\
effort & 0.679 & 0.673 & 0.651 & 0.668 & 0.760 \\
naturalness & 0.445 & 0.403 & 0.412 & 0.420 & 0.522 \\
overall & 0.646 & 0.636 & 0.607 & 0.630 & 0.732\\
\bottomrule
\end{tabular}
}
\caption{Evaluation of aspects annotated in scores using multiple annotations}
\label{tj43}
\vspace{-0.08cm}
\end{table}

\begin{table}[h]
\setlength{\tabcolsep}{3pt}
\centering
\resizebox{0.48\textwidth}{!}{
\begin{tabular}{c|ccccc}
\toprule
Low-level & \multirow{2}{*}{1} & \multirow{2}{*}{2} & \multirow{2}{*}{3} & \multirow{2}{*}{Average} & \multirow{2}{*}{Best} \\
speech aspect & & & & & \\
\midrule
noise & 0.498 & 0.499 & 0.515 & 0.504 & 0.665 \\
distortion & 0.581 & 0.580 & 0.581 & 0.581 & 0.702 \\
voice & 0.442 & 0.443 & 0.449 & 0.445 & 0.548\\
\bottomrule
\end{tabular}
}
\caption{Evaluation of aspects annotated in descriptions using multiple annotations}
\label{tj44}
\vspace{-0.08cm}
\end{table}

\end{document}